\documentstyle[pra,eqsecnum,aps,amssymb]{revtex}

\begin{document}
\draft
\title{Ground and excited states of the hydrogen negative ion in strong magnetic fields}
\author{O.-A. Al-Hujaj and P. Schmelcher}
\date{\today}
\address{Theoretische Chemie, Institut f\"ur Physikalische Chemie der Universit\"at Heidelberg, INF 229, 69120 Heidelberg, Germany}
\maketitle
\begin{abstract}
  The lowest bound states of the hydrogen negative ion and negative donor systems in a
  homogeneous magnetic field are investigated
  theoretically via a full configuration interaction approach with  an
  anisotropic Gaussian basis set. The broad  
  magnetic field regime  $\gamma=8\cdot10^{-4}-4 \cdot 10^{3}$ is
  covered. Nonrelativistic total energies,
  electron detachment energies and transition wavelengths are
  presented assuming an infinite nuclear mass. The binding
  mechanisms are  discussed in detail. The accuracy for the
  energies is  enhanced  significantly compared to 
  previously published data.
\end{abstract}
%\narrowtext

\section{Introduction}

The term ``strong field''  characterizes a situation for which  the
Lorentz force is of the order of magnitude or greater than the Coulomb binding
force. For a hydrogen atom in the ground state the corresponding field strength
cannot be reached in the laboratory, but only in astrophysical
objects like white 
dwarfs ( B$\approx$ $10^2$--$10^5$T) or neutron stars  ( B$\approx$
$10^7$--$10^9$T). Astrophysicists possess therefore a vivid interest in the
behavior and 
properties of matter in strong magnetic fields: theoretically
calculated data of magnetized atoms
can be
used for the determination of the decomposition and magnetic field configuration of
astrophysical objects \cite{Wickramasinghe:1984_1,Ruder:1994:_1,Henry:1984_1,Jordan:1998_1}.
On the other hand the strong magnetic field regime is accessible in
the laboratory if one 
considers highly excited Rydberg states of e.g. atoms
\cite{Friedrich:1989_1,Schmelcher:1998_1}.

In solid state physics donor states in semiconductors with parabolic
conduction bands are systems which possess a Hamiltonian equivalent to the
one of hydrogen within an effective mass approximation. Due
to screening effects the Coulomb force is much weaker 
than in the case of hydrogen. The regime where the ground state of the
system is dominated by
magnetic forces can therefore be reached  for certain
semiconductors in the laboratory. As an example we mention GaAs for
which  the effective mass
is $m^*=0.067~m_e$ and the static dielectric constant
$\epsilon_s=12.53~\epsilon_0$. Since the Hamiltonian of the atomic ion
and the negative donor are connected through a scaling transformation
the values for the energies given in the present work hold for both
systems equally. The reader should however keep in mind that they are given in differently
scaled units.

Apart from the above  atoms and molecules in strong magnetic fields
are also of interest  from a pure theoretical point of view. 
Due to the competition of the spherically symmetric Coulomb potential and
the cylindrically symmetric  magnetic
field interaction we encounter a nonseparable, nonintegrable problem.
 Perturbation theory, which is possible in the weak and in
the ultrastrong field regime, breaks down in the intermediate field
regime. It is therefore necessary to develop new techniques to solve such
problems. 
The neutral hydrogen atom in a strong magnetic field is now understood
to a high degree (see \cite{Friedrich:1989_1,Ruder:1994_1} and references
therein). Recently Kravchencko has published an 
``exact'' solution which provides  an infinite double sum for the eigenvalues
\cite{Kravchenko:1996_1}. With the presented method all energy values
of bound states could in principle be calculated to arbitrary precision.

For  two electron atoms the situation is significantly
different. The problems posed by the electron-electron interaction and the
 non-separability on the one-particle level have to be solved
simultaneously, which is much harder. 
The H$^-$ ion provides an additional challenge since correlation plays
an important role for its binding properties. Without a field it
possesses only one bound state \cite{Hill:1977_1}. In the presence of a magnetic
field and for the assumption of an infinitely heavy nucleus it could
be shown \cite{Avron:1981_1} that there exists an infinite number of
bound states. For laboratory field strengths these states are, due to the
binding mechanism via a one dimensional projected polarization
potential, very weakly bound \cite{Bezchastnov:unpub}. Some finite nuclear
mass effects can be included via scaling
relations\cite{Ruder:1994_1,Becken:1999_1,Pavlov-Verevkin:1980_1}. However,
the 
influence of the center of mass motion has not been investigated
in detail so far.  In the present work we 
assume an infinitely heavy nucleus which represents a good
approximation for the slow H$^-$ atomic ion in strong magnetic fields
and describes simultaneously the situation of negatively charged
donors D$^-$ in the field. Relativistic corrections were neglected
since they are assumed to be small compared to the electron detachment
energy of the system. We will  use in the following the spectroscopic
notation $^{2S+1}M$ for the electronic states of the ion where $M$ 
and $S$ are the total magnetic and spin quantum numbers.  Since
states with negative z-parity are not considered here we omit the corresponding
label in our notation (see also section \ref{subsec:sym_ham}).

Many authors have tackled the quantum
mechanical problem of H$^-$ in a strong magnetic field. One of the 
first, who pursued  a variational approach to this problem, were
Henry et al.\cite{Henry:1974_1}. They give first qualitative insights
into the  weak and intermediate field regime.  Mueller et
al. \cite{Mueller:1975} qualitatively described the
strong field ground state $^3(-1)$ and the $^10$  state  for high fields ($\gamma\approx4$ to
$\gamma\approx20\,000$, where $\gamma=1$~a.~u. corresponds to  $2.3554\cdot10^5$T).

Larsen has published a number of papers on this problem
\cite{Larsen:1979_1,Larsen:1979_2,Larsen:1981_1}. On the one hand he
created very simple and physically motivated trial functions with only
a small number of variational parameters. On the other hand his
energies were ``state of the art'' in variational calculations for a long time.
 In \cite{Larsen:1979_1} he provides  binding energies of the lowest
 $^10$ state in the  field regime $\gamma=0-5$ and of the $^3(-1)$  
state in the regime  $\gamma=0-3$. He also presents
figures showing the binding energies of the singlet and triplet
state for $M=-2$ and $M=-3$. Later \cite{Larsen:1981_1} he presents total and electron detachment energies for the lowest $^10$,
$^3(-1)$ and  $^3(-2)$
state in the high field regime. More specifically
the regime $\gamma=20-1\,000$ for the $^3(-1)$ state and $\gamma=20-200$ for the other
states were investigated.
Furthermore Park and Starace \cite{Park:1984_1} provided  upper and lower bounds for
energies and binding energies of the ground state $^10$ for weak
fields. 

In the nineties several authors
\cite{Vincke:1989_1,Larsen:1992_1,Larsen:1993_1,Blinowski:1994_1}
improved the accuracy of the binding energies 
and total energies by new techniques. Vincke and Baye
\cite{Vincke:1989_1} report total ionization energies for the 
lowest singlet and triplet states with $M=0,-1$ and $-2$
 for a few field
strengths in the regime $\gamma=4-400$. They are to our knowledge the first who
reported that the $^1(-1)$ state becomes bound for sufficiently high field strengths and
realized that the $^1(-2)$ state is slightly stronger
bound than the corresponding triplet state in the high field regime.
Larsen and McCann present in \cite{Larsen:1992_1} one-particle binding
energies for the $^10$ state  in the broad magnetic field regime  $\gamma=0-
200$. In \cite{Larsen:1993_1} the same authors consider
furthermore the singlet and triplet states of $M=-1,-2$. The
triplet states are calculated for $\gamma=0.5-200$, the
$^1(-1)$ state in the field regime $\gamma=55-2\,000$  and the
$^1(-2)$ state is calculated for a few field strength in the range
$\gamma=1-100$.
Blinowski and Szwacka \cite{Blinowski:1994_1} have subsequently used a Gaussian
basis set, similar to the 
one used in our calculation. They present results for the $^10$
state, which are less accurate than those of ref.
\cite{Larsen:1992_1}.

We also mention some Hartree--Fock calculations:
 very early Virtamo \cite{Virtamo:1976_1} has investigated the  ground
 state energies from $\gamma\approx20$ to 
$\gamma\approx20\,000$. Thurner et al.\cite{Thurner:1993_1} (results
 published in \cite{Ruder:1994_1}) have calculated triplet states
for $M$=$-1$,$-2$ and $-3$ for many field strength in the broad range
 $\gamma=2\cdot10^{-4}- 2\cdot10^{3}$. However since they use
spherical wave functions for weak  fields and cylindrical ones
for  high fields, there remains a gap of  inaccurate results in
the intermediate field regime.

In the present investigation we provide lower variational
energies and higher one-particle binding energies for the atomic H$^-$
problem and respectively the negatively charged donor center D$^-$
problem in a strong magnetic field compared to 
 all other
published data sofar. An exception is the field free
situation: the  calculation by Pekeris \cite{Pekeris:1962_1} gives
$-0.52775$~a.u. for the ground state binding energy whereas we obtain
$-0.5275488$~a.u. Clearly the field-free situation is much better understood
than the case of a strong field.

The paper is organized as follows: in section \ref{sec:sym_ham_basis}
we  consider the symmetries 
of the Hamiltonian and the basis set we  use in our
calculations. In section \ref{sec:select} we will report on the
strategy we  employed for the 
selection of basis functions in order to obtain 
accurate  results. Section \ref{sec:res_dis} contains the discussion of our 
 results and a comparison  with the literature.

\section{Hamiltonian, Symmetries and basis set}\label{sec:sym_ham_basis}

\subsection{Hamiltonian and Symmetries}\label{subsec:sym_ham}

In the following we assume an infinite nuclear mass (fixed donor). The
magnetic field is chosen to point along the z-direction.
The nonrelativistic Hamiltonian takes in atomic units 
the form 
\begin{equation}
  H=H_1+H_2+\frac{1}{\left|\bbox{r}_1-\bbox{r}_2\right|}\label{hamiltonian}
\end{equation}
with
\begin{equation}
  H_i=\frac{1}{2}\bbox{p}_i^2+\frac{1}{2}\gamma l_{z_i} + \frac{\gamma^2}{8}
  \left(x_i^2+y_i^2\right) -\frac{1}{\left|\bbox{r}_i\right|}+\gamma s_{zi}.
\end{equation}
The Hamiltonian is splitted in its one-particle operators, where
 $1/2\gamma l_{z_i}$ is the
Zeeman term, $\gamma^2/8\left(x_i^2+y_i^2\right)$ is the diamagnetic term,
$-1/\left|\bbox{r}_i\right|$ is the attractive Coulomb interaction with
the nucleus (donor) and $\gamma s_{zi}$ the spin Zeeman term (we take
 the  g-factor equal  2). The two-particle
operator $1/\left|\bbox{r}_1-\bbox{r}_2\right|$ represents the
repulsive electron-electron  interaction.

The Hamiltonian (\ref{hamiltonian}) possesses four independent
symmetries and associated quantum numbers: the total spin
$S^2$, the total z-projection of the spin $S_z$, the z-component of
the total angular momentum $M$ and the total z-parity $\Pi_z$
 (parity is also conserved but not a further independent symmetry).

\subsection{One-particle basis set}

For our calculation we use an anisotropic Gaussian basis set, which has been
put forward by Schmelcher and Cederbaum in ref. \cite{Schmelcher:1988_1},
 for the purpose of investigating
atoms and molecules in strong magnetic fields. It has already
successfully been applied to helium \cite{Becken:1999_1,Becken:1999_2}, H$_2^+$
\cite{Kappes:1996_1} and H$_2$ \cite{Detmer:1998_1}.

Adapted to the problem discussed here
this one-particle  basis set for the spatial part reads in the cylindrical
coordinates  as follows
\begin{equation}
  \Phi_i(\rho,\phi,z)=\rho^{n_{\rho_i}}z^{n_{z_i}}\; e^{-\alpha_i
  \rho^2-\beta_i z^2} \exp(i m_i \phi).
\end{equation}
These functions are eigenfunctions of the symmetry operations of the
one-particle Hamiltonian $H_i$, i.e. eigenfunctions of $l_z$
 and $\pi_z$. The additional parameters $n_{\rho_i}$ and
$n_{z_i}$ obey  the following 
restrictions:
\begin{eqnarray}
  n_{\rho_i}=|m_i|+2k_i; \qquad& k_i =0,1,2,\ldots\qquad& \textnormal{and}\quad
  m_i=\ldots,-2,-1,0,1,2,\ldots\\
  n_{z_i} =\pi_{z_i}+2l_i; \qquad& l_i =0,1,2,\ldots\qquad&
  \textnormal{and}\quad  \pi_{z_i}=0,1.
\end{eqnarray}
The exponents $\alpha_i$ and $\beta_i$ serve as positive, nonlinear variational
parameters. Due to these parameters, the one-particle functions are
flexible enough to be adapted to the situation of an arbitrary field strength: in the weak 
magnetic field regime a basis set with an almost isotropic choice of
parameters $\alpha_i\approx\beta_i$   describes the slightly  perturbed spherical symmetry.
For very high magnetic fields  it is  appropriate to choose
$\alpha=\gamma/4$ since $\rho^{|m_i|}\exp(-\gamma/4\,\rho^2)$ yields  the
$\rho$-dependence of the lowest Landau level for a given magnetic
quantum number. The $\beta_i$ will be
well tempered in a wide region. In the intermediate field regime the
basis is composed of functions with certain magnetic field dependent
sets of $\{\alpha_i,\beta_i\}$ which mediate the extreme cases.
The optimal choice is found by searching the set of $\{\alpha_i$,
$\beta_i\}$ which yields the lowest eigenvalues of the one-particle
Hamiltonian. The parameters $\{\alpha_i$,$\beta_i\}$ are successively optimized  using the pattern search
algorithm. In this manner we have optimized up to five excited states
in every symmetry subspace.
The starting values for the parameters $\{\alpha_i$,
$\beta_i\}$ have to be chosen very
carefully to find a deep local or even the global minimum. Since the search in this
high dimensional space is very time consuming,  an
optimal choice of the $k_i$ and $l_i$ is crucial: for every new
$k_i$, $l_i$ configuration a new optimization
procedure has to be started.  
The resulting binding energies for the neutral
hydrogen atom were identical to 7~--~9 digits with the one given in
\cite{Kravchenko:1996_1} for almost all field strengths for the ground
state and 5~--~7 digits were recovered for  states with higher magnetic
quantum number $|m_i|$.

We point out that Blinowski and Szwacka \cite{Blinowski:1994_1} have used a
similar basis set, but without the 
monomers $\rho^{2k_i}$ and $z^{2l_i}$. The additional
monomers however decisively enhance the flexibility and accuracy of the calculations.

\subsection{Two-particle configurations}

As a next step we build two-particle configurations from our optimized
 one-particle basis set and represent the Hamiltonian
 (\ref{hamiltonian}) in this configuration space. This is done for each
 total symmetry $(S^2,\Pi_z,L_z)$ separately. The corresponding
 spectrum of H$^-$ is then obtained by diagonalizing the Hamiltonian matrix.
 We hereby use all possible excited two-particle configurations
 constructed from
 our optimized one-particle basis set, i.e. our
approach is a full configuration interaction method (full CI). The
two-particle functions are   
constructed from the one-particle functions by selecting 
combinations for $m_i+m_j=M$ and $\pi_{zi}+\pi_{zj}=\Pi_z$. The spin
part can be trivially separated. 
Due to the antisymmetrization of the spatial wave function the configuration
space of the triplet states is  slightly smaller than 
that of the singlet states since for  triplet configurations
there are no combinations with $i=j$.

As our basis set is not orthogonal we have to solve a
finite-dimensional generalized real symmetric eigenvalue problem
\begin{equation}
  (\underline{\underline{H}}-E\underline{\underline{S}})\cdot\underline{c}=0
\end{equation}
where $\underline{\underline{H}}$ is the matrix representation of the
Hamiltonian  and $\underline{\underline{S}}$ the overlap matrix. The
resulting energies $E$ are strict upper bounds to the exact
eigenvalues  in the given
subspace of symmetries. 

Some technical remarks concerning the calculation of the matrix
elements are in order. All matrix elements can be evaluated
analytically. With the exception of the electron-electron integrals
all expressions can be calculated very rapidly. The electron-electron
integrals, however, deserve a special treatment: through a combination
of transformation techniques as well as analytical continuation
formulae for the series of involved transcendental functions their
representation has been simplified enormously (for details see
ref. \cite{Becken:1999_1} and in particular \cite{Becken:1999_2}). It is due to this extremely efficient
implementation of the electron-electron integral that  large basis
sets of the order of $2500-4000$ could be used in the present work  to perform CI calculations
for many field strengths.

\section{Selection of the basis functions}\label{sec:select}

Since the single bound state in the absence of the external field is bound only 
due to  correlation, and all the other states in the presence of the
magnetic field are only weakly bound, it is
very important to include correlation by a proper choice of the
one-particle basis functions building up the two-particle configurations.  
For the $M=0$ singlet state this was achieved by selecting
one-particle basis functions not only with $m_1=m_2=0$ but also with 
$m_1=-m_2\not=0$. This allows one to describe the angular correlation
which is particular important for the $^10$ state. In general the
enhanced binding properties of  negative ions in the presence of a
magnetic field are due to a balanced competition of the different
interactions. On the one hand the confinement due to the magnetic
field raises the kinetic energy and the electrostatic repulsion due to
the electron-electron interaction. These effects tend to lower the
binding energy. On the other hand the confinement raises the nuclear
attraction energy, the exchange energy and to some extent also the
correlation energy which tend to enhance the binding energy. Of course
one has to distinguish between, for example, the $^10$ state whose
binding properties are dominated by correlation effects and the
excited bound states with nonzero magnetic quantum numbers which
possess a significant contribution to their binding energy through exchange
effects and due to the occupation of the series of tightly bound
hydrogenic orbitals  $1s$, $2p_{-1}$, $3d_{-2}$,\ldots etc. %On

For the description of the lowest states with $|M|>0$ an effective
one-particle picture can be employed  \cite{Larsen:1979_1}: the
hydrogen negative ion consists of a tightly bound core electron with
magnetic quantum number zero and a significantly less bound electron which
carries 
the magnetic quantum number of the ion.
The core electron is then described  by one-particle basis functions with
 $m_1=0$. The outer electron is described by  one-particle
functions with $m_2=M$ in order to take into account the fact  that it
is weakly bound and thus spatially 
extended. 
In order to go beyond this effective one-particle picture we used,
similar to the case $M=0$ one-particle functions with
other magnetic quantum numbers to obtain in particular the correlation behavior.

The above picture is not valid for the tightly bound states in the 
high field regime: the number of
functions with different magnetic quantum numbers can be reduced as we increase the field
strength. This reduction in the
number of basis functions is also suggested by the occurrence of  linear
dependencies for strong fields. The extent
of this reduction can be seen from the fact that the number of  two-particle
basis functions drops from $4\,000$ for $\gamma=0$ to less than $3\,000$
for $\gamma=4000$ for the $^10$ state.

\section{Results and discussion}\label{sec:res_dis}

As already mentioned the H$^-$ ion possesses only one bound state in
the absence of the magnetic field \cite{Hill:1977_1}. Turning on the field it has been
shown \cite{Avron:1981_1} that there exists (for infinite nuclear
mass) for any nonzero field  an infinite number of bound states. The
corresponding proof \cite{Avron:1981_1} relies on the physical
picture \cite{Bezchastnov:unpub} that the external electron  is for
weak fields far from the neutral atomic core and experiences
therefore to lowest order  a polarization potential due to the
induced dipole moment of the core. Perpendicular to the field the
motion of the external electron is dominated by the field and it
occupies approximately Landau orbitals whereas parallel to the field
it is weakly bound due to the projection of the mentioned
polarization potential on the Landau orbitals which yields an
one-dimensional binding along the field. For typical strong laboratory fields
the corresponding binding energies are of the order of $10^{-6}$~eV for the
hydrogen atom negative ion and significantly larger for more electron atoms
with a larger polarizability.
 To investigate theses states in the weak field regime
goes clearly beyond the feasibility of the present method. Instead we
will investigate a number of states, starting from the value of the
field strength for which they become significantly bound, which means that
the outer electron is already relatively close to the core and possesses a
binding energy of at least a few meV. Clearly in that case the picture
of the polarization potential is no more valid since exchange and
correlation effects rule the binding properties of the ion. Within
our approach we could find one bound state for each negative magnetic quantum
number of 
the ion considered ($-3\leq M\leq 0$) for both singlet and triplet
states, except the $^30$ state, which is unbound. Their behavior has
been studied for the complete range of 
field strengths $0.01\leq\gamma\leq4000$.
 The one bound state of the
H$^-$ ion in the absence of the field represents, in the above sense, an
exception since it is already significantly bound without the field.
All these states possess positive z-parity and \emph{no bound states
  could be found for negative z-parity}.

\subsection{Threshold energies}

The electron detachment energy is defined to be the energy we need to remove
one electron from the atom
without changing the quantum numbers of the total system.
The corresponding lowest threshold energy $E_T$ for the
H$^-$ ion  can be expressed as:
\begin{equation}
  E_T=\frac{\gamma}{2}\left(|M|+M+2+g_e M_s\right)-I(\textnormal{H})
\end{equation}
where $I($H$)$ is the binding energy of the ground state of the neutral hydrogen
atom in a magnetic field. The term 
$\gamma/2(|M|+M+2)$ is the energy of an electron in the
lowest Landau level with magnetic quantum number $m=M$ where the
spin part is omitted. This means that the free electron carries  the whole
angular momentum of the state. For magnetic quantum numbers
$M\leq0$ the threshold energy $E_T$ is independent of the angular momentum $M$, i.e. there
is only a singlet and a triplet threshold.  The threshold energy is then
$E_T=\gamma-I($H$)$ for singlet states and  $E_T=-I($H$)$ for triplet
states. 
We denote the electron detachment energy by $I(H^-)$ which is
 given by 
 $I($H$^-)=E_T-E_{tot}$
where $E_{tot}$ is the total energy of the considered state of H$^-$.

\subsection{Total,  electron detachment and transition energies}

Before we discuss the individual states and their  properties let us describe
 some general features of the states 
considered here. The total energy of the singlet states is monotonically
increasing with increasing field strength. This fact is caused by the 
increase of the field-dependent kinetic energy.
 In contrast to this the 
total energy of the triplet states is monotonically decreasing with
increasing field strengths. This is a consequence of the  additional
 spin Zeeman term (we
consider here only the $S_z=-1$ component of the spin triplet states).
{\em The electron detachment energies are monotonically increasing with
increasing  field strength for all states considered here, i.e. both
singlet and triplet states.\/} This has to be seen in view of the above-mentioned fact 
that the zero-point kinetic (Landau) energy of the electrons is raised in the presence of the
magnetic field and therefore the threshold energy for loosing one electron
is raised in the same way.

For the $^10$ state the total energy raises from $-0.52754875$ at
$\gamma=0$ to $3986.49870$ at $\gamma=4000$. This state is the most tightly
bound state for all field strengths. The detachment energy
increases from $0.027549$ a.u. at $\gamma=0$ to $2.29805$ a.u. at
$\gamma=4000$. There are two reason which give rise to the fact, that
this state is the most tightly bound one. On the one hand the
electrons are in this state much closer to the nucleus than in other
states. This increases the binding  due to the attractive nuclear potential energy. On
the other hand  correlation has an  important impact on the binding energy. Both
effects are reinforced with increasing  field strength as the
electrons become more an more confined in the x-y plane perpendicular to the
magnetic field. These effects overcome the influence of the
static electron-electron repulsion.
The total energies and the detachment
energies of the $^10$ state are presented in table
\ref{tab:0Sbindcomp}. It can be seen that the detachment energies for
this most tightly bound state could be improved by 1-2\% for 
all field strengths compared to the existing literature.
This is not correct for a vanishing field, where much more efficient
basis sets like the Hylleraas basis set are available.
For numerical reasons the relative accuracy for the detachment
energies is largest in the intermediate field regime.

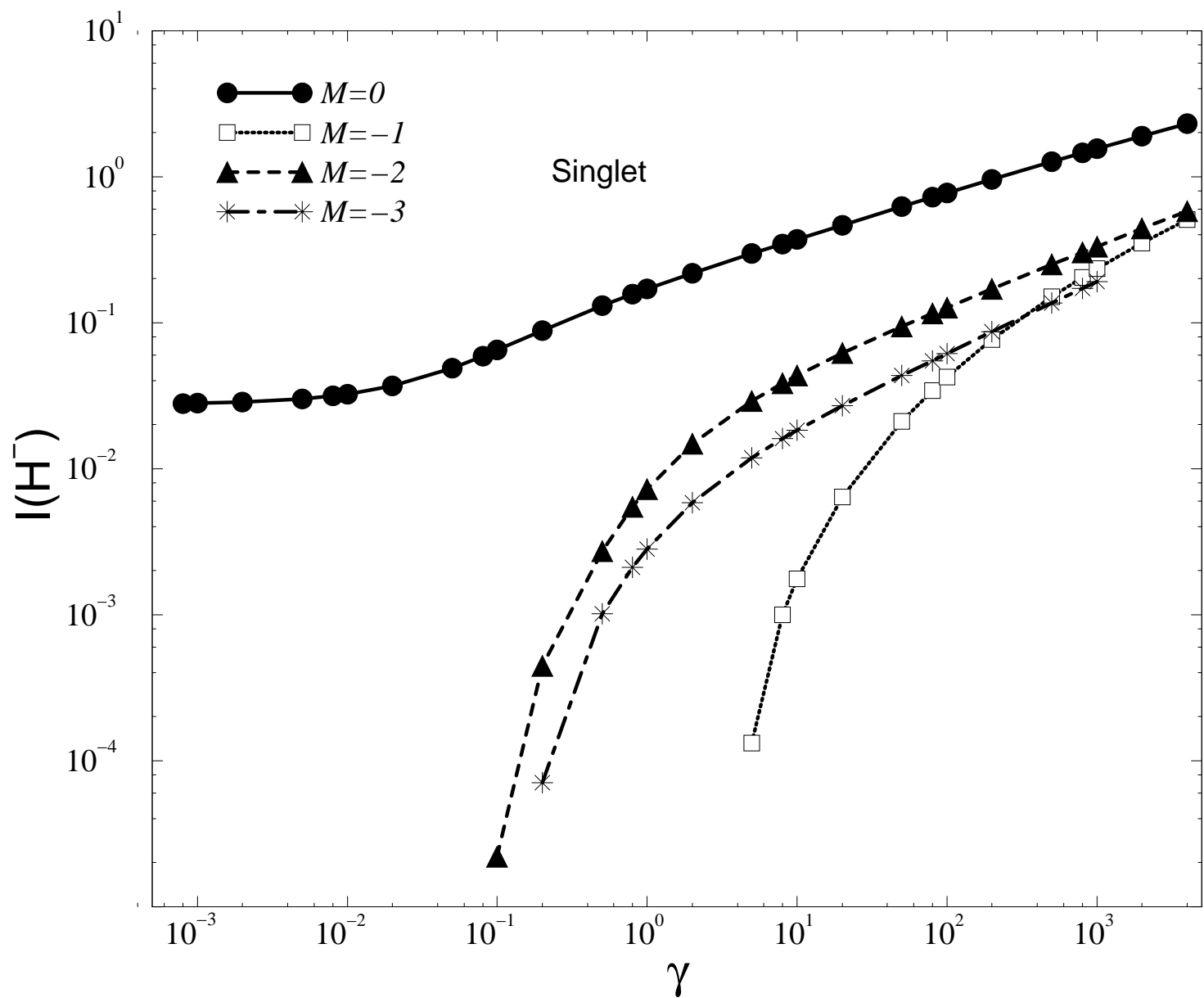
\begin{figure}[htbp] 
  \caption{Electron detachment energies of the singlet states in
    atomic units.
     Note that the $M=-1$ singlet state is not bound for weak fields
     and is for $5\lesssim\gamma\lesssim300$ weaker bound than all
     other states considered. For $\gamma\approx300$ and
     $\gamma\gtrsim4000$ we encounter crossings of the detachment
     energies of the $^1(-1)$ with those of the $^1(-3)$ and $^1(-2)$
     states, respectively.
 \label{fig:sinbind}}
\end{figure}

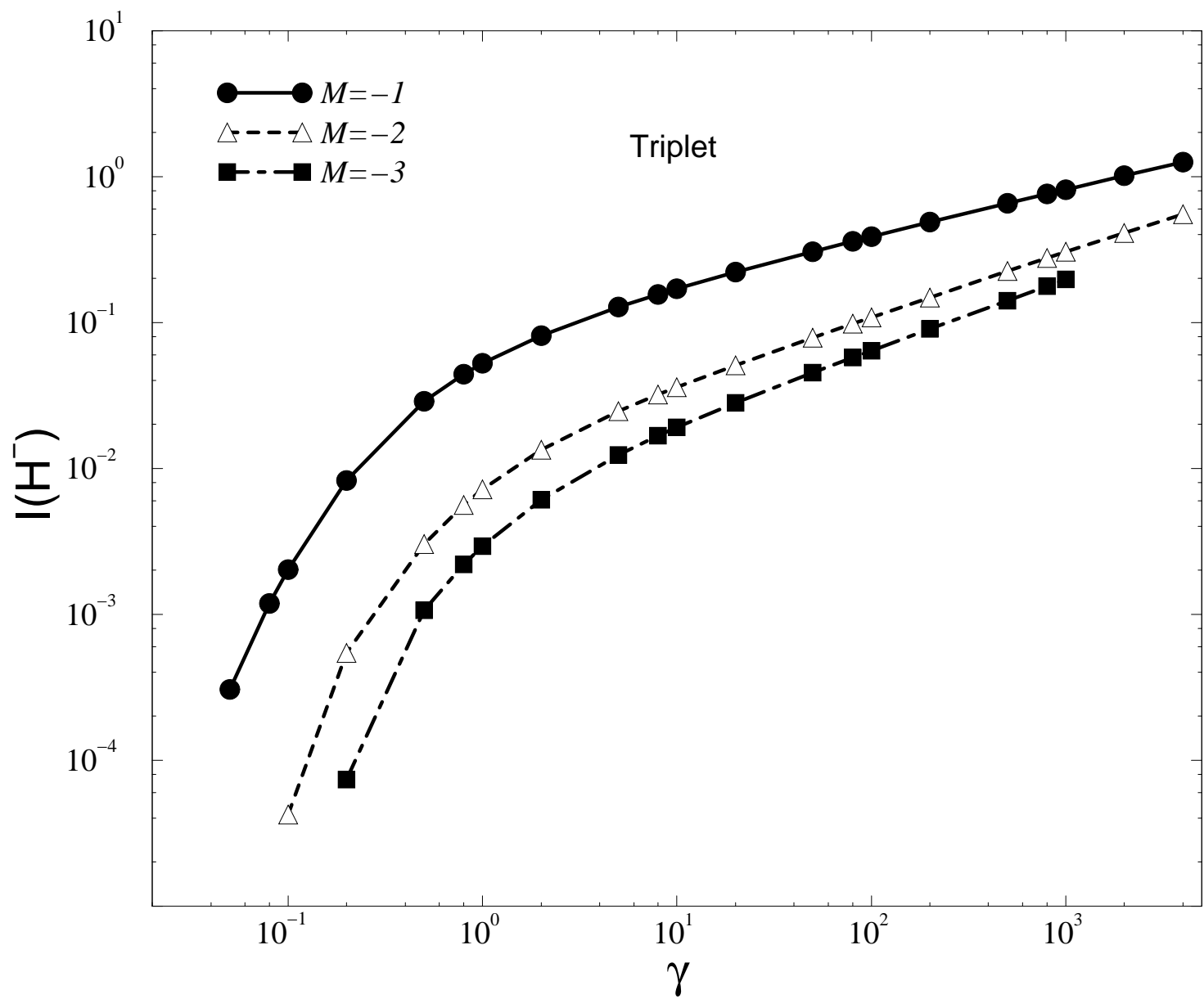
\begin{figure}[htbp]
  \caption{Electron detachment energies of the triplet states in
    atomic units. 
    \label{fig:tripbind}}
\end{figure}

The $^30$ state is not bound for all considered field strengths. 
This can be understood in an effective particle picture as follows:
for triplet states the spatial two-particle wave function is
antisymmetric with respect to particle exchange and therefore the two
particles have to occupy different spatial orbitals, i.e. we are
exclusively dealing with excited configurations. For $M\not=0$ it is
(see later) possible to obtained tightly bound triplet states in a
strong magnetic field by occupying different orbitals of the
hydrogenic series ($1s, 2p_{-1}, 3d_{-2}, \ldots$) which yields the
one-particle excited configurations of the type $1s2p_{-1},
1s3d_{-2}, \ldots$. For the case of the $^30$ state, however, we have
$M=0$ and only configurations constructed from pairs of two orbitals
with $(m,-m)$ are allowed which are either of doubly excited character
($m\not=0$) or a singly excited configuration with $m=0$. Therefore no
magnetically tightly bound configurations are allowed for the $^30$
state which illuminates its unbound character for any field
strength. 
All singlet and triplet electron detachment
energies of all the considered bound states are  presented
also graphically:
 Figure \ref{fig:sinbind} shows the singlet detachment energies
and figure \ref{fig:tripbind}  the corresponding energies for the
triplet states. 

It is important to mention that the global ground state of the ion undergoes
a crossover with respect to its symmetry with increasing field
strength. For weak fields the $^10$ state is the 
ground state of the system, whereas in strong fields the $^3(-1)$ state
becomes the ground state which was first shown in ref.\cite{Henry:1974_1}.
This is caused by the spin
Zeeman term, which lowers the total energy of the triplet states. The 
crossover takes place at $\gamma_c\approx0.05$ which corresponds to
approximately $10^4$~T for the H$^-$ ion. The $^3(-1)$ state is very weakly bound when
it becomes the ground state (at $\gamma_c$ the detachment
energy is $\approx 3\cdot10^{-4}$ a.u.). This  prevents us from
 localizing more exactly the field strength at which the crossover takes place.
The $^3(-1)$ state, being the ground state of the anion for $\gamma>\gamma_c$ never
becomes the most tightly bound state.  At $\gamma=4\,000$ its
electron detachment energy is $1.25$ a.u. and therefore much less than
the detachment energy of the $^10$ state. 
This is due to the fact that the tightly bound states are formed by
occupying the hydrogenic series $1s, 2p_{-1}, 3d_{-2}, \ldots$ (as
mentioned above) and the $^10$ states allows for the $1s^2$
configuration yielding the strongest binding although it represents an
excited state for $\gamma > \gamma_c$ due to its spin character.

The
singlet state $^1(-1)$  is not bound for weak fields. It becomes bound
in the regime
$\gamma\approx1-5$ which is an unexpected behavior. The $^1(-1)$ state
lies  higher
in the spectrum than the bound $^1(-2)$ and $^1(-3)$ states for
the intermediate field region. In the high field region it
however crosses both
states. The crossing with the $^1(-3)$
takes place at $\gamma\approx300$, the crossing with the $^1(-2)$
state is at $\gamma\gtrsim4\,000$. Unfortunately the accuracy of our
method is not sufficient to provide a closer look at this crossing.
The fact that the $^1(-1)$ state is not bound for weak fields but bound for
strong fields is a consequence of the complicated interplay of the
different interactions.
 The Coulomb repulsion of the two electrons is much
weaker for the spatially antisymmetric triplet states compared to the singlet states.
The electron-electron  repulsion is higher for the states
with $M=-1$ compared to the states with $M<-1$.
This pushes the $|M|=1$ singlet states for weak
fields beyond the threshold energy, i.e. makes them unbound.
The total ionization and the detachment energies of the singlet and
triplet states with $M=-1$ are 
presented in table \ref{tab:singtrip1}. 
The suppression of the binding for the singlet state can clearly be
 seen from this table:
 the detachment energy of the singlet  is $100$ times  lower
 than for the triplet at $\gamma=10$,
 but at $\gamma=4000$ the ratio is of the order $2$.
The comparison with the literature (see table \ref{tab:singtrip1})
shows that our detachment energies are variationally lower by several
percent than the best available data. For the situation of weakly
bound states the improvement is significantly larger.

Let us now consider the energies for the states with $M=-2$ which are presented in table
\ref{tab:singtrip2}.
Focusing on the detachment energies we realize that
for weak fields the triplet  state possesses a larger detachment energy than the
singlet state, but for intermediate and high fields the singlet
state is stronger bound than the triplet one, i.e. we encounter a
crossover which is presented in figure  \ref{fig:bindcross2s2t}. 
Compared to the data of ref.\cite{Larsen:1993_1}, our method yields $5-10$\% higher
variational detachment energies for the triplet state and several times
higher detachment energies for the singlet one.
 If we consider the
singlet-triplet splitting which is the  difference of the total
energies between the
singlet and the triplet state, where the  spin-Zeeman shift is
omitted, it can be observed, that for all states this splitting behaves
monotonically increasing with increasing field strength in the weak
field regime. The splitting
for the states with $M=-2$ and $M=-3$ are shown in figure
\ref{fig:singtripsplitee}.  The splitting
for the $M=-2$ states   increases in weak fields, but for
high fields this splitting decreases and
becomes negative above some critial field strength. It seems that the
Coulomb repulsion, due to 
antisymmetrization of the wave function is dominated by correlation
effects. 
That the above observation is in fact a consequence of correlation is
supported by  Vincke and Baye \cite{Vincke:1989_1}: the reversed
order concerning the detachment energies (see figure
\ref{fig:bindcross2s2t}) occurs if
they  include so-called transverse mixing, which  simulates
correlations in their approach. 

For states with $M=-3$  only a few
published data are available. These states are only weakly bound, although they
are stronger bound for $\gamma\gtrsim300$ than the $^1(-1)$ state. The singlet
state has for $\gamma=0.2$ a detachment energy of
$7.1~10^{-5}$~a.u.\ and  at $\gamma=1000$ its detachment energy is
 $0.19$. The electron detachment energies of the
triplet state are of the same order of magnitude and the absolute value of
the singlet triplet splitting is the lowest of the 
states considered here. As a consequence a careful convergence study
of the results (detachment energy) is indispensable.  Our
data are given in table \ref{tab:singtrip3}. 

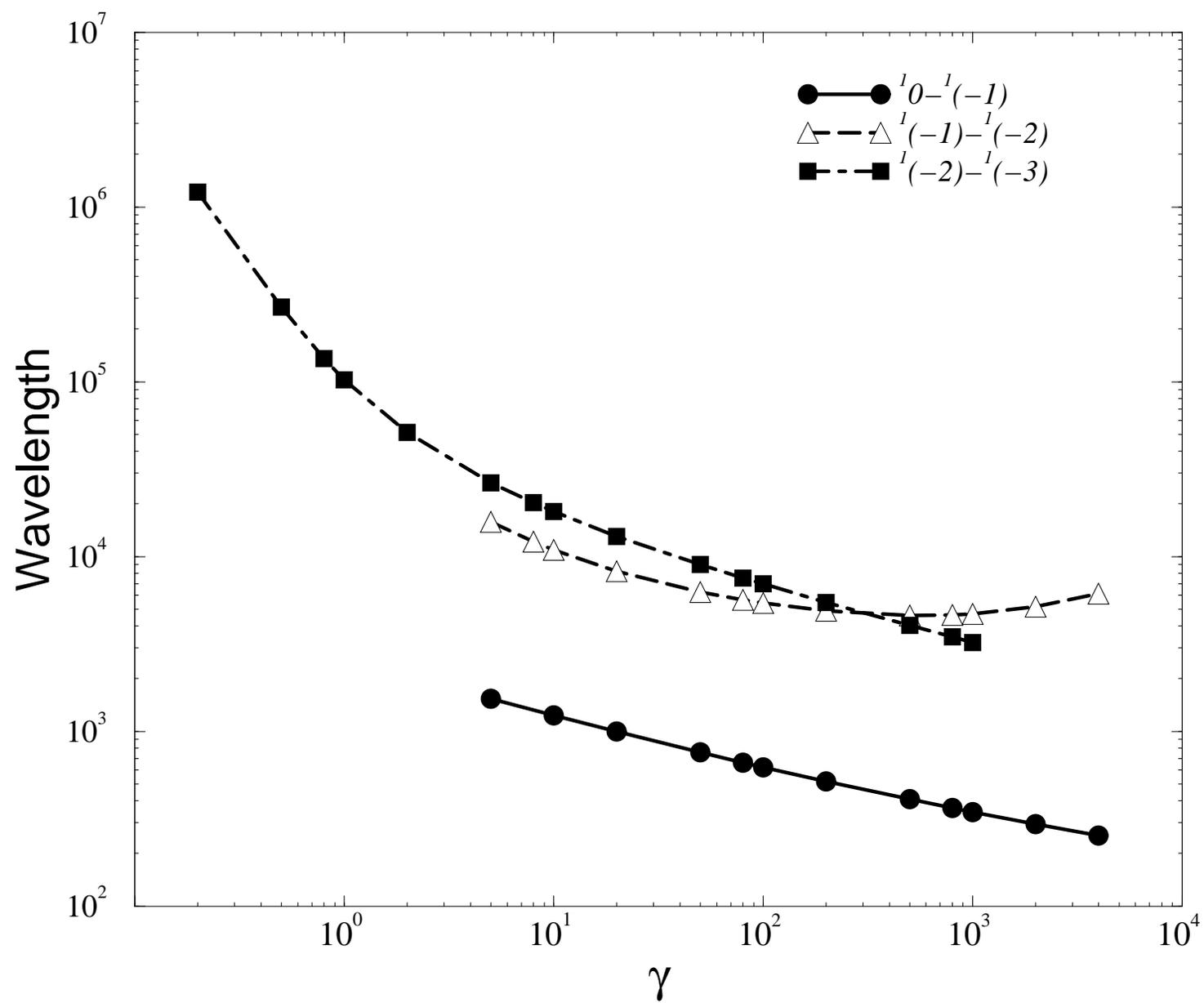
\begin{figure}[!htbp]
  \caption{Singlet transition wavelengths between the considered {\em bound} states in
    {\AA}ngstr{\o}m as a function of the field strengths in atomic
  units on a logarithmic scale.}
\label{fig:singtrans}
\end{figure}

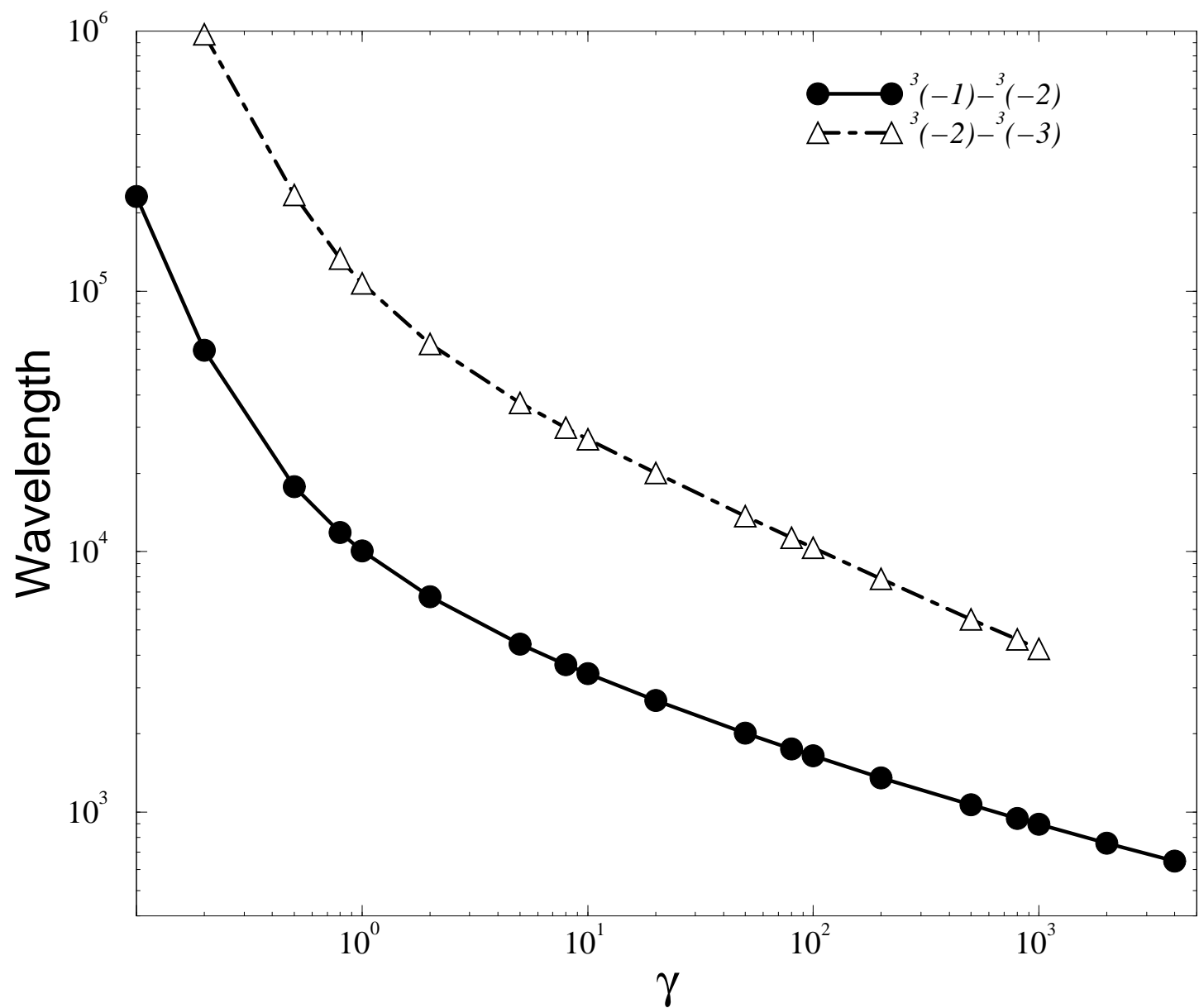
\begin{figure}[!htbp]
  \caption{Triplet transition wavelengths between the considered states in
    {\AA}ngstr{\o}m as a function of the field strength in atomic units on a logarithmic scale.}
  \label{fig:triptrans}
\end{figure}

The wavelengths of the transitions of the singlet states are presented in figure
\ref{fig:singtrans}. The wavelengths are monotonically decreasing with
incresing field strength except
for the transition from the $^1(-1)$ state to the $^1(-2)$ state. As mentioned above these
states cross at $\gamma\gtrsim4\,000$. Therefore the corresponding wavelength for
this transition  diverges at the crossing field strength. 
The  transition wavelengths for the triplet states shown in figure
\ref{fig:triptrans} are also monotonically
decreasing with increasing field strength. 

Finally we comment on corrections due to the finite nuclear mass. There are
two kinds of corrections, which are relevant here. One, which is
special for ions in strong magnetic fields and which describes the
coupling between the center of mass motion and the electronic
motion. 
This coupling is due to a motional electric field of intrinsic
dynamical origin seen by the moving ion in a magnetic field \cite{Schmelcher:1991}. Second
there are corrections due to the replacement of the naked masses by
reduced ones which can be easily included in our data by performing
the corresponding shifts
\cite{Becken:1999_1,Becken:1999_2,Pavlov-Verevkin:1980_1} . 
A full
dynamical treatment of the atomic ion including the collective motion
goes clearly beyond the scope of the present investigation. It is
important to note that for the case of the fixed negative donors there
naturally occur no such corrections.

\section{Brief Summary}

We have investigated the H$^-$ ion, negative donors D$^-$
respectively,  in a strong magnetic field via a fully 
correlated approach. The key ingredient is an
anisotropic Gaussian basis set, whose one-particle wave functions
are nonlinearly optimized in order to obtain the spectrum of the
one-particle Hamiltonian. In
contrast to other basis sets, which are appropriate either for the low
field or for the high field regime, our basis set is flexible enough
to be adapted to the situation of arbitrary field strength and
especially suited for the intermediate field regime. All calculations
were performed in the infinite mass frame neglecting relativistic corrections.

We have investigated the low field ground state $^10$, as well as
singlet and triplet states for $M=-1,-2,-3$ for the broad field regime
$\gamma=8\cdot 10^{-4}-4\cdot10^{3}$\@. For all states and almost all field
strengths we could reach at least $1-2\%$ higher binding energies,
compared to all other published data. For some states our binding energies were
larger by a factor up to two.  The global ground state undergoes a 
crossover with respect to its symmetry which is well-known in the literature \cite{Henry:1974_1}:
for weak fields $\gamma\lesssim5\cdot10^{-2}$ the
global ground state is the $^10$ state, whereas for
$\gamma\gtrsim5\cdot10^{-2}$ it is the $^3(-1)$ state, which is much
weaker bound than the $^10$ state for all field strengths. 
The $^1(-1)$ state becomes bound for $\gamma\lesssim5$
and it crosses the $^1(-3)$ state at $\gamma\approx300$ and the
$^1(-2)$ state at $\gamma\gtrsim4000$. We have also investigated the
electronic states with $M=-2$ in detail. For $\gamma\lesssim1$ the triplet state
is stronger bound than the singlet, whereas for $\gamma\gtrsim1$ the
singlet is stronger bound than the triplet. Explanations for the
binding mechanisms of the considered states have been provided. The transition wavelengths for all
allowed transitions as a function of the field strength are thereby obtained. No
stationary transitions which could be of relevance to the astrophysical observation
in magnetized white dwarfs have been observed.

\subsection*{Acknowledgments}

The Deutsche Forschungsgemeinschaft (OAA) is gratefully acknowledged
for financial support. We thank W. Becken for many fruitful
discussions and for his help concerning computational aspects of
the present work.

\newpage

\begin{center}
{\bf{\large Tables}}
\end{center}

\vspace*{3.0cm}

\narrowtext
\narrowtext
\squeezetable
\begin{table}[!htbp]

\caption{Nonrelativistic fixed nucleus total and electron detachment energies of the
  field free ground state $^10$ ($^1S_0$) of H$^-$. We also provide the results
  for the electron detachment energies given in the literature so
  far.\label{tab:0Sbindcomp}} 
\begin{tabular}{dddd}
  $\gamma$      & E$_{tot}$    & I(H$^-$)& I$_{Lit}$(H$^-$)       \\
\tableline                                                        
    0           &-0.52754875   &0.02754875  & 0.2775\tablenotemark[1]\\   
    8 10$^{-4}$ &-0.52754430   &0.02794446  &                        \\  
    1 10$^{-3}$ &-0.52754053   &0.02804078  &0.02735\tablenotemark[2]\\
    2 10$^{-3}$ &-0.52753777   &0.02853877  &0.02785\tablenotemark[2]\\
    5 10$^{-3}$ &-0.52749800   &0.03000425  &0.0293 \tablenotemark[2]\\
    8 10$^{-3}$ &-0.52740873   &0.03142473  &                        \\
    0.01        &-0.52734972   &0.03237472  & 0.0317\tablenotemark[2]\\
    0.02        &-0.52677018   &0.03687014  & 0.0362\tablenotemark[2]\\
    0.05        &-0.52314046   &0.04876375  &                        \\
    0.08        &-0.51715770   &0.05874669  &                        \\
    0.1         &-0.51223522   &0.06470874  & 0.0634\tablenotemark[3]\\
    0.2         &-0.47868356   &0.08830200  &0.08685\tablenotemark[3]\\
    0.5         &-0.32804874   &0.13083820  & 0.130\tablenotemark[4] \\
    0.8         &-0.13939006   &0.15710667  &                        \\
     1.0        &-0.00178881   &0.17061922  & 0.1695\tablenotemark[4]\\
    2.0         & 0.75990486   &0.21788123  & 0.2175\tablenotemark[4]\\
     5.0        & 3.3234387    &0.2961625   &0.2955\tablenotemark[4] \\
     8.0        & 6.0350437    &0.3455713   &                        \\
    10.0        & 7.8806402    &0.3715626   & 0.371\tablenotemark[4] \\   
    20.0        & 17.319887    &0.464715    & 0.463\tablenotemark[4] \\   
    50.0        & 46.359385    &0.622747    & 0.618\tablenotemark[4] \\   
    80.0        & 75.753770    &0.721953    &                        \\   
   100.0        & 95.436219    &0.773977    &0.7665\tablenotemark[4] \\   
   200.0        & 194.31374    &0.95911     &0.9385\tablenotemark[4] \\   
   500.0        & 492.47687    &1.26604     &                        \\   
   800.0        & 791.35985    &1.45501     &                        \\    
   1000.0       & 990.78459    &1.55299     &                        \\   
   2000.0       &1988.8003     &1.8949      &                        \\   
   4000.0       &3986.4978     &2.2981      &                        \\   
                                                                      
\end{tabular}                                                         
\tablenotetext[1]{See Pekeris \cite{Pekeris:1962_1}.}
\tablenotetext[2]{See Park et al.\cite{Park:1984_1}.}
\tablenotetext[3]{See Larsen \cite{Larsen:1979_1}.}
\tablenotetext[4]{See Larsen \cite{Larsen:1992_1}.}

\end{table}

\narrowtext
\widetext
\begin{table}[!htbp]
%\squeezetable
  \caption{Nonrelativistic total and electron detachment energies
    (atomic units) of singlet and triplet states with $M=-1$ 
    as a function of the magnetic field strength $\gamma$. These
    states evolve from the $^1P_{-1}$ and $^3P_{-1}$ states for zero
    magnetic field. \label{tab:singtrip1}} 
  \begin{tabular}{dddd ddd}
          &    \multicolumn{3}{c }{$^1(-1)$}                &  \multicolumn{3}{c}{$^3(-1)$}                            \\
  $\gamma$&  E$_{tot}$ & I(H$^-$) & I$_{Lit}$(H$^-$)         & E$_{tot}$      & I(H$^-$) & I$_{Lit}$(H$^-$)               \\
  \tableline                                                 
    0.05  &            &          &                          &   -0.52468218  & 0.00030547      &0.00025\tablenotemark[1]  \\
    0.08  &            &          &                          &   -0.53959063  & 0.00117963      &                          \\
    0.1   &            &          &                          &   -0.54954554  & 0.00201906      & 0.0016\tablenotemark[1]  \\
    0.2   &            &          &                          &   -0.59861960  & 0.00823804      & 0.0072\tablenotemark[1]  \\
    0.5   &            &          &                          &   -0.72586763  & 0.02865709      &0.027875\tablenotemark[2] \\
    0.8   &            &          &                          &   -0.82643425  & 0.04415086      &                          \\  
     1.0  &            &          &                          &   -0.88359474  & 0.05242585      & 0.0518\tablenotemark[2]  \\
    2.0   &            &          &                          &   -1.1036308   & 0.0814168       & 0.0805\tablenotemark[2]  \\
     5.0  &  3.6194699 &0.0001312 &                          &   -1.5081497   & 0.1277508       & 0.1263\tablenotemark[2]  \\
     8.0  &  6.3792043 &0.0009946 &                          &   -1.7751617   & 0.1557768       &                          \\ 
    10.0  &  8.2504454 &0.0017574 &                          &   -1.9181202   & 0.1703230       & 0.168\tablenotemark[2]   \\
    20.0  &  17.778209 &0.006392  &                          &   -2.436716    & 0.221318        & 0.2175\tablenotemark[2]  \\
    50.0  &  46.961089 &0.021050  &                          &   -3.323515    & 0.305655        & 0.309\tablenotemark[2]   \\
    80.0  &  76.441490 &0.034306  &                          &   -3.882658    & 0.358381        &                          \\
   100.0  &  96.167783 &0.042402  & 0.00281\tablenotemark[2] &   -4.175890    & 0.386100        & 0.38015\tablenotemark[2] \\
   200.0  &  195.19141 &0.07671   & 0.0407 \tablenotemark[2] &   -5.21214     & 0.48500         &  0.4771\tablenotemark[2] \\
   500.0  & 493.59274  &0.15017   &                          &   -6.90934     & 0.65226         &                          \\
   800.0  & 792.61061  &0.20426   &                          &   -7.94296     & 0.75783         &                          \\
   1000.0 & 992.10306  &0.23452   & 0.1727 \tablenotemark[2] &   -8.47584     & 0.81343         &                          \\
   2000.0 &1990.3441   &0.3511    & 0.2732 \tablenotemark[2] &   -10.3165     & 1.0117          &                          \\
   4000.0 &3988.2901   &0.5057    &                          &   -12.4576     & 1.2534          &                          \\
\end{tabular}                                                    
\tablenotetext[1]{See Larsen \cite{Larsen:1979_1}}
\tablenotetext[2]{See Larsen and McCann \cite{Larsen:1993_1}.}
\end{table}

\widetext
\begin{table}[!htbp]
%\squeezetable
  \caption{Nonrelativistic total eigenenergies and electron detachment
    energies (atomic units) of singlet and triplet states with $M=-2$ 
    as a function of the magnetic field strength $\gamma$. These
    states evolve from the $^1D_{-2}$ and $^3D_{-2}$ states in the
    absence of a magnetic field. \label{tab:singtrip2}} 
  \begin{tabular}{dddd ddd}
          &    \multicolumn{3}{c }{$^1(-2)$}                &  \multicolumn{3}{c}{$^3(-2)$}                            \\
  $\gamma$&  E$_{tot}$ & I(H$^-$)    & I$_{Lit}$(H$^-$)        & E$_{tot}$      & I(H$^-$)       & I$_{Lit}$(H$^-$) \\
  \tableline                                                 
    0.1   &  -0.44754846 &2.20 $10^{-5}$&                          &  -0.54756898   & 4.25 $10^{-5}$ &                         \\
    0.2   &  -0.39082732 &0.00044576     &                         &  -0.59092467   &0.00054311    &                         \\
    0.5   &  -0.19993316 &0.00272262     &                         &  -0.70022629   &0.00301575    & 0.0023\tablenotemark[1] \\
    0.8   &   0.01224680 &0.00546981     &                         &  -0.78790031   &0.00561691    &                         \\ 
     1.0  &   0.16158323 &0.00724788     &  0.0015\tablenotemark[1]&  -0.83835741   &0.00718852    & 0.0064\tablenotemark[1] \\
    2.0   &   0.96305720 &0.01472889     &                         &  -1.03557327   &0.01335936    & 0.0123\tablenotemark[1] \\
     5.0  &   3.5905353  &0.0290659      &                         &  -1.4049975    &0.0245987     & 0.02290\tablenotemark[1]\\
     8.0  &   6.3421331  &0.0384819      &                         &  -1.6512176    &0.0318326     &                         \\
    10.0  &   8.2087000  &0.0435028      &  0.006\tablenotemark[1] &  -1.7838114    &0.0360142     & 0.0335\tablenotemark[1] \\
    20.0  &  17.722831   &0.061762       &                         &  -2.2663464    &0.0509479     & 0.047 \tablenotemark[1] \\
    50.0  &  46.888223   &0.093916       &                         &  -3.096448     &0.078587      & 0.0719\tablenotemark[1] \\
    80.0  &  76.360491   &0.115232       &                         &  -3.621917     &0.097640      &                         \\
   100.0  &  96.083393   &0.126803       &  0.03\tablenotemark[1]  &  -3.897965     &0.108128      & 0.09895\tablenotemark[1]\\
   200.0  & 195.10291    &0.16995        &                         &  -4.87531      &0.14817       & 0.13535\tablenotemark[1]\\
   500.0  & 493.49377    &0.24914        &                         &  -6.48097      &0.22388       &                         \\
   800.0  & 792.51251    &0.30235        &                         &  -7.46090      &0.27576       &                         \\
   1000.0 & 992.00594    &0.33164        &                         &  -7.96662      &0.30420       &                         \\
   2000.0 &1990.2556     &0.4397         &                         &  -9.7152       &0.4105        &                         \\
   4000.0 &3988.2159     &0.5799         &                         & -11.7535       &0.5493        &                         \\
\end{tabular}                                                    
\tablenotetext[1]{See Larsen and McCann \cite{Larsen:1993_1}.} 
\end{table}

\widetext
\begin{table}[!htbp]
%\squeezetable
  \caption{Nonrelativistic total  and electron detachment energies
    (atomic units) of singlet and triplet states with $M=-3$ 
    as a function of the magnetic field strength $\gamma$. These
    states evolve from the $^1F_{-3}$ and $^3F_{-3}$ states in the
    absence of a magnetic field. \label{tab:singtrip3}} 
  \begin{tabular}{ddd ddd}
          &    \multicolumn{2}{c }{$^1(-3)$}                       &  \multicolumn{3}{c}{$^3(-3)$}                     \\
  $\gamma$&  E$_{tot}$   & I(H$^-$)            & E$_{tot}$         & E$_{tot}$ (Lit.)          & I(H$^-$)              \\
  \tableline                                                 
    0.2   &   -0.39045232&7.075$\cdot 10^{-5} $& -0.59046132       &                           & 7.975$\cdot 10^{-5} $ \\
    0.5   &   -0.19822524& 0.00101470          & -0.69827772       &                           & 0.00106718             \\
    0.8   &    0.01561745& 0.00209916          & -0.38448370       &                           & 0.00220031             \\
     1.0  &    0.16602786& 0.00280324          & -0.83410348       &-0.7092618\tablenotemark[1]& 0.00293459             \\
    2.0   &    0.97196637& 0.00581972          & -1.02829096       &                           & 0.00607705             \\
     5.0  &    3.6077963 & 0.0118049           & -1.3927524        &                           & 0.0123535              \\
     8.0  &    6.3646091 & 0.0160059           & -1.6361379        &                           & 0.0167528              \\ 
    10.0  &    8.2339720 & 0.0182308           & -1.7669036        &                           & 0.0191064              \\
    20.0  &   17.757749  & 0.026853            & -2.243571         &                           & 0.028172               \\
    50.0  &   46.938916  & 0.043223            & -3.063189         &                           & 0.045328               \\
    80.0  &   76.420911  & 0.054812            & -3.581678         &                           & 0.057401               \\
   100.0  &   96.148914  & 0.061282            & -3.853929         &                           & 0.064125               \\
   200.0  &  195.18642   & 0.08643             & -4.81732          &                           & 0.09017               \\
   500.0  &  493.60717   & 0.13574             & -6.39799          &                           & 0.14091               \\
   800.0  &  792.64412   & 0.17074             & -7.36188          &                           & 0.17675               \\
  1000.0  &  992.14738   & 0.19020             &  -7.8591          & -7.686295\tablenotemark[1]& 0.19663               \\
\end{tabular}                                                      
\tablenotetext[1]{See Thurner et al. \cite{Ruder:1994_1}.}
\end{table}
                             
\narrowtext

\widetext

\newpage
\begin{center}
{\bf{\large Figure Captions}}
\end{center}

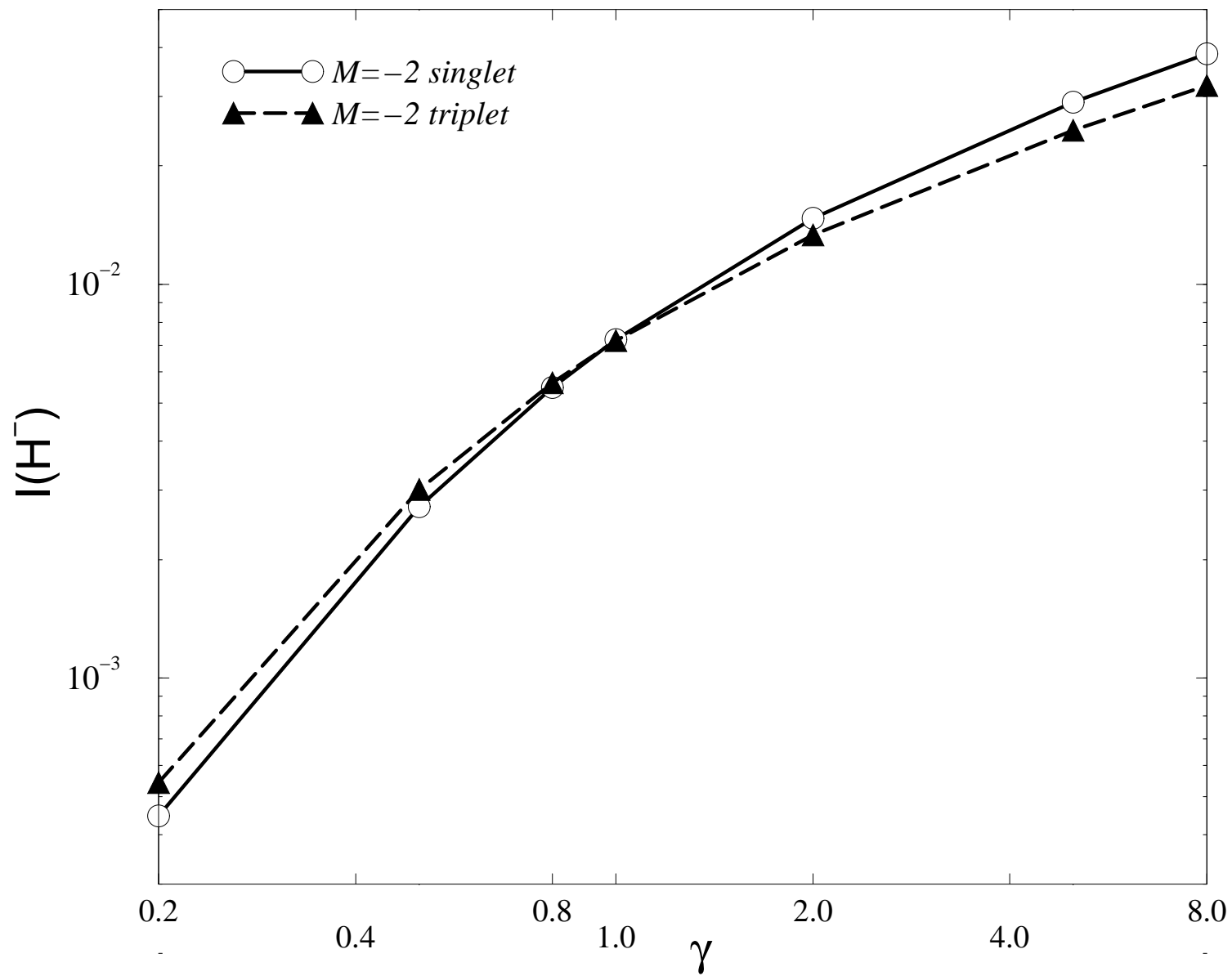
\begin{figure}[htbp]
  \caption{Electron detachment energies of the singlet and triplet
    state of the $M=-2$ states in atomic units.  Note that the singlet 
  and triplet state reverse their order: For low field strengths the
  triplet state is more bound whereas for high fields ($\gamma\gtrsim 1$) the singlet state is more bound than the
  triplet state.\label{fig:bindcross2s2t}}

\end{figure}
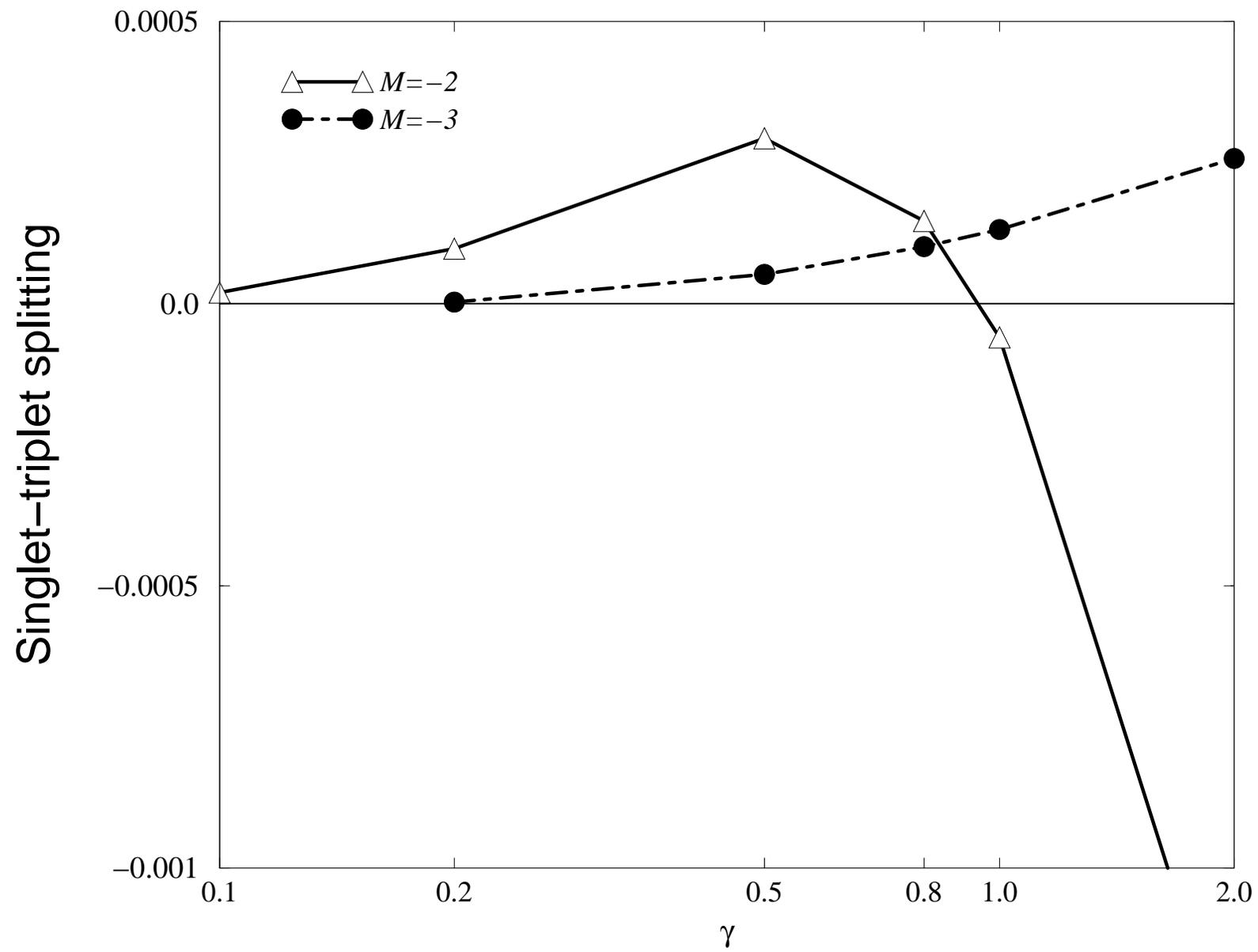
\begin{figure}[htbp]
  \caption{Singlet triplet splitting for the $M=-2$ and the $M=-3$
    states in atomic units. The splitting due to the spin Zeeman term
   is omitted. Note that the splitting of the 
   $M=-2$ states increases for low fields  but decreases for high
   fields, whereas the splitting for the $M=-3$ states 
   increases monotonically with increasing
   field strength.}
  \label{fig:singtripsplitee}
\end{figure}

\begin{thebibliography}{10}

\bibitem{Wickramasinghe:1984_1}
D. Wickramasinghe and L. Ferrario, Astrophys. J. {\bf 282},  222  (1984).

\bibitem{Ruder:1994:_1}
H. Ruder, G. Wunner, H. Herold, and F. Geyer, {\em Atoms in strong magnetic
  fields} (Springer Verlag, Berlin, 1994).

\bibitem{Henry:1984_1}
R. Henry and R. O'Connell, Astrophys. J. {\bf 282},  L97  (1984).

\bibitem{Jordan:1998_1}
S. Jordan, P. Schmelcher, W. Becken, and W. Schweizer, Astron. Astrophys. {\bf
  336},  L33  (1998).

\bibitem{Friedrich:1989_1}
H. Friedrich and D. Wintgen, Phys. Rep. {\bf 183},  37  (1989).

\bibitem{Schmelcher:1998_1}
{\em Atoms and Molecules in Strong External Fields}, edited by P. Schmelcher
  and W. Schweizer (Plenum Press, New York, 1998).

\bibitem{Ruder:1994_1}
H. Ruder, G. Wunner, H. Herold, and F. Geyer, {\em Atoms in Strong Magnetic
  Fields} (Springer, Heidelberg, 1994).

\bibitem{Kravchenko:1996_1}
Y.~P. Kravchenko, M.~A. Liberman, and B. Johansson, Phys. Rev. A {\bf 54},  287
   (1996).

\bibitem{Hill:1977_1}
R.~N. Hill, Phys. Rev. Lett. {\bf 38},  643  (1977).

\bibitem{Avron:1981_1}
J.~E. Avron, I.~W. Herbst, and B. Simon, Commun. Math. Phys. {\bf 79},  525
  (1981).

\bibitem{Bezchastnov:unpub}
V. Bezchastnov, P. Schmelcher, and L.~S. Cederbaum, 1999, subm. to Phys. Rev.
  A.

\bibitem{Becken:1999_1}
W. Becken, P. Schmelcher, and F.~K. Diakonos, J. Phys. B {\bf 32},  1557
  (1999).

\bibitem{Becken:1999_2}
W. Becken and P. Schmelcher, subm. to J. Phys. B.

\bibitem{Pavlov-Verevkin:1980_1}
V. Pavlov-Verevkin and B.~I. Zhilinskii, Phys. Lett. A {\bf 78 A},  244
  (1980).

\bibitem{Henry:1974_1}
R.~J. Henry, R.~F. {O'Connel}, and E. Smith, Phys. Rev. D {\bf 9},  329
  (1974).

\bibitem{Mueller:1975}
R.~O. Mueller, A. Rau, and L. Spruch, Phys. Rev. A. {\bf 11},  789  (1975).

\bibitem{Larsen:1979_1}
D.~M. Larsen, Phys. Rev. B {\bf 20},  5217  (1979).

\bibitem{Larsen:1979_2}
D.~M. Larsen, Phys. Rev. Lett. {\bf 42},  742  (1979).

\bibitem{Larsen:1981_1}
D.~M. Larsen, Phys. Rev. B {\bf 23},  4076  (1981).

\bibitem{Park:1984_1}
C.-H. Park and A.~F. Starace, Phys. Rev. A {\bf 29},  442  (1984).

\bibitem{Vincke:1989_1}
M. Vincke and D. Baye, J. Phys. B {\bf 22},  2089  (1989).

\bibitem{Larsen:1992_1}
D.~M. Larsen and S.~Y. McCann, Phys. Rev. B {\bf 46},  3966  (1992).

\bibitem{Larsen:1993_1}
D.~M. Larsen and S.~Y. McCann, Phys. Rev. B {\bf 47},  13175  (1993).

\bibitem{Blinowski:1994_1}
J. Blinowski and T. Szwacka, Phys. Rev. B {\bf 49},  10231  (1994).

\bibitem{Virtamo:1976_1}
J. Virtamo, J. Phys. B {\bf 9},  751  (1976).

\bibitem{Thurner:1993_1}
G. Thurner {\it et~al.}, J. Phys. B {\bf 26},  4719  (1993).

\bibitem{Pekeris:1962_1}
C.~L. Pekeris, Phys. Rev. {\bf 126},  1470  (1962).

\bibitem{Schmelcher:1988_1}
P. Schmelcher and L.~S. Cederbaum, Phys. Rev. A {\bf 37},  672  (1988).

\bibitem{Kappes:1996_1}
U. Kappes and P. Schmelcher, Phys. Rev. A  {\bf 51},
  4542 (1995); {\bf 54},  1313  (1996); Phys. Lett. A {\bf 210}, 409 (1996).

\bibitem{Detmer:1998_1}
T. Detmer, P. Schmelcher, and L.~S. Cederbaum, Phys. Rev. A {\bf 56},  1825
  (1997); {\bf 57}, 1767 (1998).

\bibitem{Schmelcher:1991}
P. Schmelcher and L.~S. Cederbaum, Phys.Rev. A 43, 287 (1991)

\end{thebibliography}
\end{document}